\documentclass[12pt,preprint]{aastex}
\slugcomment{Accepted for the publication in the {\it Astrophysical Journal}}
\shortauthors{Mori}
\shorttitle{X-ray Spectral Analysis of The Crab Nebula}

\begin{document}
\title{Spatial Variation of the X-ray spectrum of the Crab Nebula}

\author{Koji Mori and David N.\ Burrows}
\affil{Department of Astronomy and Astrophysics, Pennsylvania State
University, 525 Davey Laboratory, University Park, PA. 16802, USA}
\email{mori@astro.psu.edu}

\author{J.\ Jeff Hester}
\affil{Department of Physics and Astronomy, Box 871504, Arizona State
University, Tyler Mall, Tempa, AZ. 85287-1504, USA}

\author{George G.\ Pavlov}
\affil{Department of Astronomy and Astrophysics, Pennsylvania State
University, 525 Davey Laboratory, University Park, PA. 16802, USA}

\author{Shinpei Shibata}
\affil{Department of Physics, Yamagata University, Yamagata
990-8560, Japan}

\and
\author{Hiroshi Tsunemi}
\affil{Department of Earth and Space Science, Graduate School of
        Science, Osaka University, 1-1 Machikaneyama, Toyonaka, Osaka
        560-0043, Japan}

\begin{abstract}

We present spectral analysis of the Crab Nebula obtained with the {\it
Chandra} X-ray observatory.  The X-ray spectrum is characterized by a
power-law whose index varies across the nebula. The variation can be
discussed in terms of the particle injection from the pulsar in two
different directions: the equatorial plane containing the torus and the
symmetry axis along the jet. In the equatorial plane, spectra within the
torus are the hardest, with photon index $\alpha \approx 1.9$, and are
almost independent of the surface brightness. At the periphery of the
torus, the spectrum gradually softens in the outer, lower surface
brightness regions, up to $\alpha \approx 3.0$. This indicates that
synchrotron losses become significant to X-ray emitting particles at the
outer boundary of the torus. We discuss the nature of the torus,
incorporating information from observations at other wavelengths.
Spectral variations are also seen within the southern jet.  The core of
the jet is the hardest with $\alpha \approx$ 2.0, and the outer sheath
surrounding the core becomes softer with $\alpha$ up to 2.5 at the
outermost part.  Based on the similarity between the spectra of the jet
core and the torus, we suggest that the electron spectra of the
particles injected from the pulsar are also similar in these two
different directions.  The brightness ratio between the near and far
sides of the torus can be explained by Doppler boosting and relativistic
aberration; however, the observed ratio cannot be derived from the
standard weakly magnetized pulsar wind model. We also found a site where
an optical filament comprised of supernova ejecta is absorbing the soft
X-ray emission ($<$ 2 keV).

\end{abstract}


\section {\label {sec:intro} INTRODUCTION}

The observed pulsating radiation is only a small fraction of a
pulsar's spin-down energy. The bulk of this energy is carried off in
the form of a magnetized, relativistic pulsar wind. This pulsar wind
is randomized by a shock where its pressure balances that of the
ambient material (Rees \& Gunn 1974; Kennel \& Coroniti 1984). Diffuse
synchrotron radiation is expected to arise from the randomized pulsar
wind, forming a {\it pulsar wind nebula} (PWN). Although this general
idea is widely accepted, details of the formation are still
controversial. For example, it is a long-standing problem that the
ratio of the Poynting flux to the particle kinetic flux just upstream
of the shock, $\sigma$, is much smaller than 10$^{-2}$, in spite of
expectations that the ratio at the light cylinder is larger than unity
(e.g., Arons 2002). Additionally, the formation of the torus and the
collimation of the pulsar wind to form a jet along the rotational axis
are quite curious but unresolved problems.

An X-ray observation is a powerful probe to peer into the inner PWN,
whose spectral and morphological properties provide essential
information to answer such questions. Recent {\it Chandra} and {\it
XMM-Newton} observations increased the number of objects where
spatially resolved X-ray spectroscopy within the PWN can be
performed. In a number of PWNe, the spectrum of the outer edges of the
PWN is observed to be softer than the spectrum of the center (e.g.,
G21.5$-$0.9, Slane et al.\ 2000, Safi-Harb et al.\ 2001; 3C 58,
Bocchino et al.\ 2001; G0.9$+$0.1, Porquet et al.\ 2003; G54.1$+$0.3,
Lu et al.\ 2002; Vela, Kargaltsev \& Pavlov 2004). Since the
synchrotron energy losses are proportional to the square of the particle
energy, the higher energy particles have shorter life times and the
spectrum becomes softer as they move down the post-shock flow, which
is in general agreement with these observations. 
Such spatial variation of the spectrum can be used to
verify theoretical models and to provide new diagnostics for the study
of PWNe.

The Crab Nebula has been one of the best examples for investigating a
PWN. Its X-ray brightness and large apparent structure including the
torus (Aschenbach \& Brinkmann 1975) and jets (Brinkmann, Aschenbach, \&
Langmeier 1985) make it an ideal target for spatially resolved X-ray
spectroscopy.  Previous {\it Chandra} and {\it XMM-Newton} observations
have also shown spectral softening of the Crab (Weisskopf et al.\ 2000;
Willingale et al.\ 2001). Here, we present spectral analysis results
from multiple observations of the Crab Nebula with the Advanced CCD
Imaging Spectrometer (ACIS) on board {\it Chandra}, which are derived
from the monitoring program described by Hester et al.\ (2002).

\section{\label{sec:obs} OBSERVATION AND ANALYSIS}

Owing to the X-ray brilliance of the Crab Nebula, observations with
the {\it Chandra} ACIS instrument are technically challenging. The high
count rate generally causes two problems in the data acquisition:
telemetry saturation and event pile-up (caused by more than one photon
depositing charge in a given pixel during a CCD exposure, thereby
compromising our ability to measure the energy of each individual
photon). Although telemetry saturation results in loss of observing
efficiency, it does not lead to biased estimates of physical
parameters (e.g., count rate) because events are discarded in a unit
of frames, not event by event. On the other hand, pile-up makes
interpretation of spectral results problematic, since it converts two
photons into a single ``event'' which has apparent energy equal to the
sum of the photon energies.  This results in underestimated count rate
and hardens the detected spectrum, as discussed in Weisskopf et al.\
(2000).

\begin{figure}
\epsscale{1.0}
\plotone{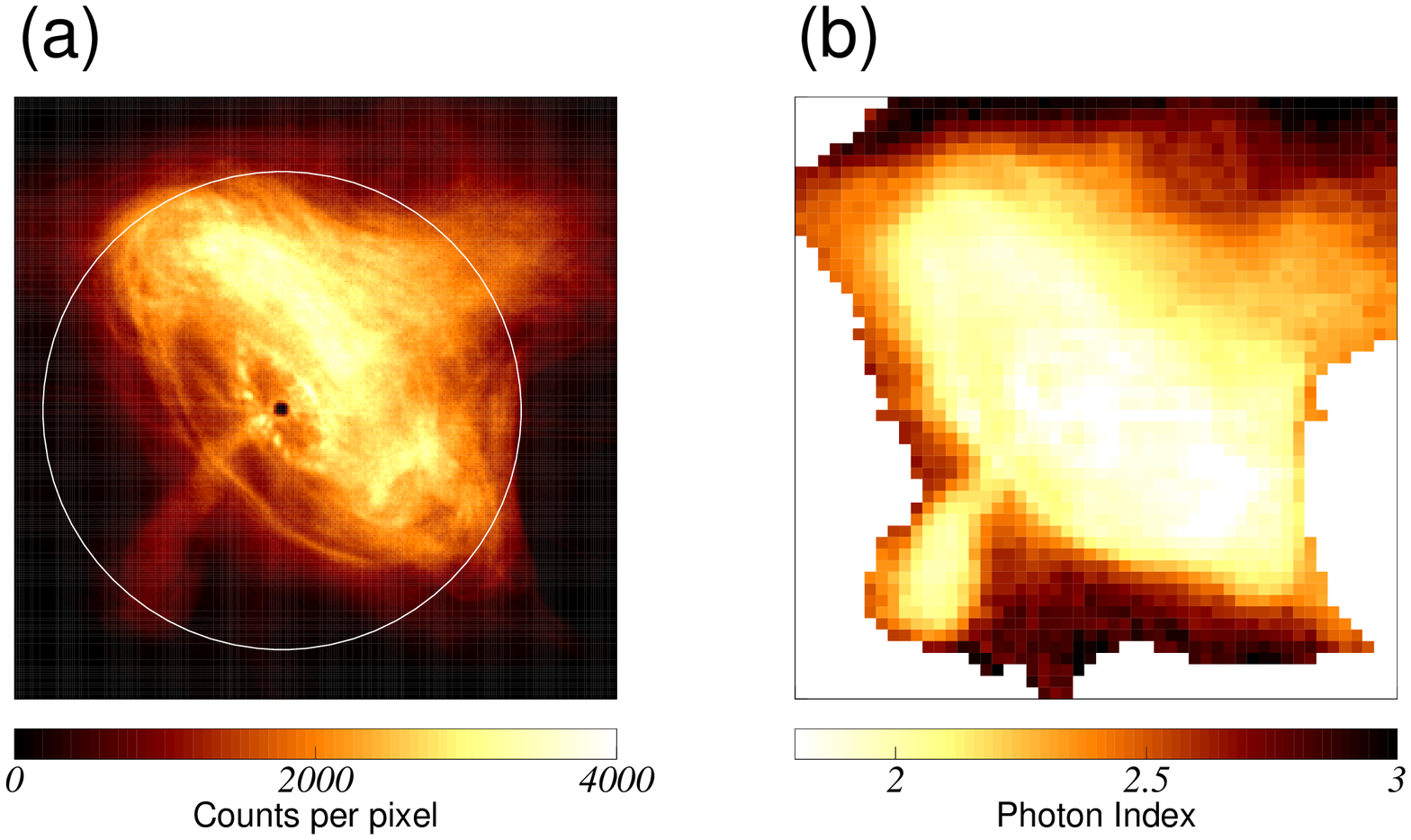}
\caption{({\it a}) {\it Chandra} ACIS-S combined image
of the 2nd$-$8th observations. 
The center of the white circle ($R=50''$) is located
at the pulsar position. The ``hole'' at the pulsar position is
caused by severe event pile-up resulting in the rejection of
most events at this position.  Narrow lines through the pulsar are 
instrumental artifacts due to trailing events.  
({\it b}) Photon index map of the Crab Nebula
after correction for pile-up effects. White regions around the eastern,
western, and southern edges were excluded from analysis due to the
dominance of trailing events and scattered photons. The pulsar
position is also white because no photon index could be derived there
due to severe pile-up.
\label{fig:index_map}}
\end{figure}

We observed the Crab Nebula eight times (Table~\ref{tbl:log}), using
an instrumental mode of ACIS-S different from that of Weisskopf et
al.\ (2000) in order to reduce pile-up. We adopted a frame time of 0.2
sec instead of the standard frame time of 3.2 seconds, but we did not
utilize the grating to reduce count rate. This reduced the counts
pixel$^{-1}$ frame$^{-1}$ by a factor of 2.3 with respect to the data
of Weisskopf et al.\ (2000), while improving statistics by increasing
the total number of counts by one order of magnitude for a similar
effective exposure time.  Since the Crab Nebula is a diffuse source,
the event pile-up was suppressed by more than a factor of 2.3.  The
observational parameters (e.g., telemetry format of graded mode, no
dithering) were the same through the eight observations. A combined
image from the last seven observations is shown in
Figure~\ref{fig:index_map}a (the first observation did not include the
whole remnant due to an instrument setup error).  This image is an
order of magnitude deeper than the image of Weisskopf et al.\ (2000).

In order to study X-ray spectral variations within the Crab Nebula, we
divided the whole image into $2.^{\!\!\prime\prime}5 \times
2.^{\!\!\prime\prime}5$ square regions, from which spectra were
extracted.  The eight data sets were combined in the analysis in order
to provide each region with enough statistics. We excluded regions
dominated by scattered photons (less than 2000 counts per analysis
region) and/or by trailing (or ``out of time'') events. Trailing
events, which are caused by photons detected during the CCD readout,
can be seen not only as the line emanating from the pulsar along the
detector readout direction (which is almost east-west for all of our
observations), but also as diffuse emission which dominates the faint
regions to the east and west of the nebula due to the enormous
brilliance of the Crab PWN. The number of the trailing events in a
given pixel can be estimated by multiplying the total number of events
integrated along the readout direction by the ratio of the charge
transfer row period to the frame exposure time (this ratio is $1.89
\times 10^{-4}$). We exclude regions where more than 20\% of the total
events recorded in that region are due to the trailing events.  

\begin{figure}
\epsscale{0.9}
\plotone{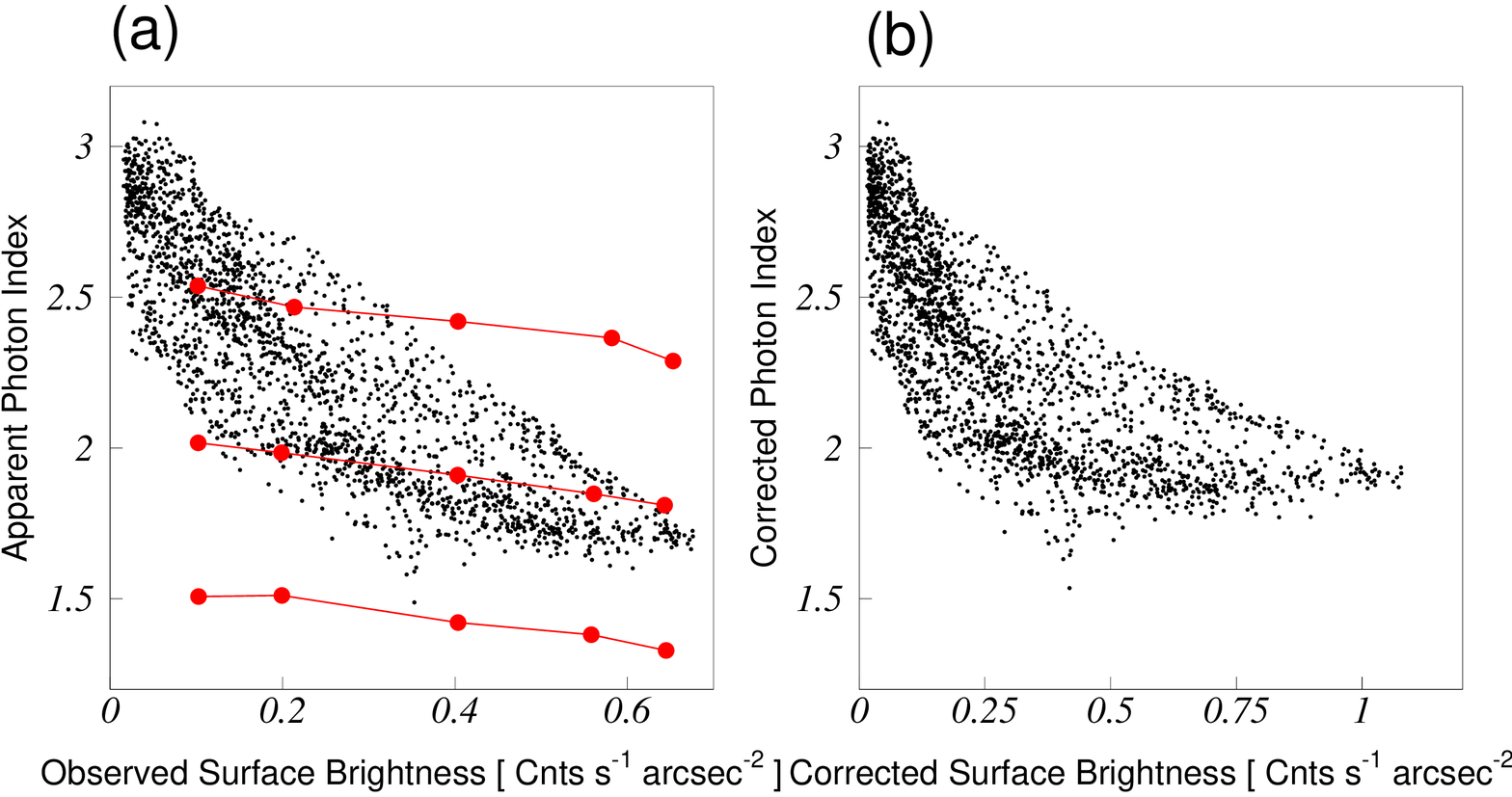}
\caption{({\it a}) 
Plot of apparent photon index against observed surface brightness
derived from $2.^{\!\!\prime\prime}5 \times 2.^{\!\!\prime\prime}5$
square regions. Results of pile-up simulations for incident photon
index of 1.5, 2.0, and 2.5 are superimposed (red lines), which shows
that pile-up results in smaller photon index in higher surface
brightness. ({\it b}) Same as (a) but corrected for pile-up effects.
\label{fig:correlation_plot}}
\end{figure}

We fitted the 2074 spectra with a power-law model in the 0.5--8 keV
energy band. We included galactic absorption with a fixed $N_{\rm H}$
of $3.2 \times 10^{21} $cm$^{-2}$, which is derived from a fit of the
faintest region where pile-up is negligible. The recently discovered
quantum efficiency decay of ACIS\footnote{See
http://asc.harvard.edu/cal/Acis/Cal\_prods/qeDeg/index.html} is taken
into account\footnote{See
http://www.astro.psu.edu/users/chartas/xcontdir/xcont.html}.
Figure~\ref{fig:correlation_plot}a shows a correlation plot of
apparent photon index versus observed surface brightness.  The
statistical errors of the apparent photon indices are quite small.
They can be roughly expressed empirically as $0.01 \times I^{-0.5}$
where $I$ is the observed surface brightness in unit of counts
s$^{-1}$ arcsecond$^{-2}$ (90\% confidence level).

In spite of the reduction in counts per frame, many of the spectra are
still distorted by event pile-up. The spectral parameters derived from
a fit of the whole nebula spectrum with a power-law plus galactic
absorption model are summarized in Table~\ref{tbl:whole_fit} (uncorrected values), 
as are the canonical values for comparison. The pile-up results in a lower
apparent spectral index and lower apparent intensity than the
canonical values. The observed intensity normalization is about 84\%
of that estimated from the canonical values with the instrumental
response.  Pile-up has the effect of reducing photon index in regions
of increased surface brightness, which is the same general trend seen
in the data of Figure~\ref{fig:correlation_plot}a.  The pile-up
effects must be corrected in order to determine the nature of any real
spectral variations across the Crab nebula.

In order to estimate how pile-up distorts the incident spectrum, we
utilized the spectral simulator tool LYNX (Chartas et al.\ 2000). LYNX
traces the propagation of individual photons through the mirror and
ACIS via the raytrace tool MARX (Wise et al.\ 1997) and the PSU ACIS
Monte Carlo CCD simulator (Townsley et al.\ 2001), respectively. Since
LYNX takes into account the possible overlap of the resulting charge
clouds within each CCD exposure frame, spectra affected by pile-up can
be simulated for both point sources and diffuse sources. We simulated
the observed Crab spectrum for several different values of true
surface brightness and for true photon indices of 1.5, 2.0, and 2.5.
The results are superposed on the data in
Figure~\ref{fig:correlation_plot}a.  The expected reduction of
apparent photon index in higher surface brightness regions is
evident. However, it is also clear that the variations in observed
photon indices greatly exceed those expected from pile-up, and there
are certainly real spectral variations across the Crab Nebula, as
described by Weisskopf et al.\ (2000).

We used the LYNX simulation results to correct the data points in
Figure~\ref{fig:correlation_plot}a for the photon index and surface
brightness errors induced by the pile-up effects.  For each data
point, we interpolated the simulated data points to the observed
surface brightness and photon index and calculated the corresponding
true surface brightness and photon index.
Figure~\ref{fig:correlation_plot}b presents a correlation plot of the
corrected data points. This simulation was intended to give
corrections on the data points in Figure~\ref{fig:correlation_plot}a
and was not adjusted to correct the whole nebula spectrum in order to
match to the canonical spectral parameters. 
However, we used the corrected photon indices
and surface brightnesses of each data point in
Figure~\ref{fig:correlation_plot}b to calculate a corrected spectrum
for the whole nebula and we present the results in Table~\ref{tbl:whole_fit}
(corrected values). The photon index and the
normalization factor of the corrected spectrum agree well with the
canonical values within the errors, which strongly supports the
success of our pile-up corrections.  Hereafter, we use the terms ``photon
index'' and ``surface brightness'' to mean these corrected values.

We should note a limitation of our simulation. Our corrections are
based on modeling of a uniform diffuse source. We apply them to each
$2.^{\!\!\prime\prime}5 \times 2.^{\!\!\prime\prime}5$ square region
under the reasonable assumption that the surface brightness within
each region is uniform. Therefore, application of this technique to
small complicated structures like the inner ring or the
relativistically expanding wisps requires more complex modeling which
is beyond the scope of this work. The systematic errors of the photon
index mainly come from this assumption.  Judging from the results of
the simulation shown in Figure~\ref{fig:correlation_plot}a and typical
surface brightness fluctuations of $\sim 10\%$ within an analysis
region, the systematic errors in photon index are $\sim$ 0.05 and
dominate the statistical errors.

Figure~\ref{fig:index_map}b shows a map of the photon index. The
structure in the torus is more symmetrical about the pulsar in the
photon index map than in the broad-band image
(Fig.~\ref{fig:index_map}a).  While Doppler boosting and relativistic
aberration brighten the northwestern portion of the torus (Pelling et
al.\ 1987) in the Figure~\ref{fig:index_map}a image, by contrast there
is relatively little variation of the photon index within the
torus. The hardest structures in the nebula, with photons indices as
low as 1.8, are the inner ring and portions of the torus, including
the circular structures seen at each extremity of the torus discovered
in the first {\it Chandra} observation (Weisskopf et al.\
2000). However, these regions contain small complex structures and the
hardest photon index should be re-examined using data set free from
pile-up.  The entire torus is quite hard ($\alpha \approx 1.9$)
compared with the outer portions of the nebula.  The southern jet is
also relatively hard ($\alpha \approx 2.0$), whereas the northern
counter jet is significantly softer ($\alpha \approx 2.25$). The
bright region around the counter jet to the northwest of the torus,
hereafter called the ``umbrella'' because of its shape, has even
softer emission ($\alpha
\approx 2.5$).  Photon indices as large as 3.0 are found in the outer
peripheral portions of the nebula.

\section{\label{sec:analysis} RESULTS}

We studied the spatial variation of the photon index over the nebula.
As a first step, we divided the nebula into four prominent regions:
the torus, the umbrella, the jet, and the peripheral region. We took
the torus region as an ellipse bounding the far side of the torus. We
also added two circles at each extremity of the ellipse to include the
circular structures in the torus region. The rest of the bright nebula
was taken as the umbrella region. The jet region covers only the
southern jet. The peripheral region includes the faint emission
surrounding the torus and umbrella regions except for the
jet. Figure~\ref{fig:color_plot} shows that those four regions occupy
separate regions in the plot of photon index against surface
brightness, with the umbrella region significantly overlapping the
torus and periphery. The color coding is shown in the inset figure.

The data points of the region within the torus (red) form a band with
nearly constant photon index of $\approx 1.9$ over a wide range of
surface brightness (more than a factor of 4). Although the spectral
index increases slightly for surface brightness below 0.4 counts
s$^{-1}$ arcsecond$^{-2}$, this result shows that our spectral
corrections were successful in removing the spectral distortion due to
pile-up.

The data points of the peripheral region (green) dominate the low
surface brightness / high photon index region. The photon indices at
the periphery of the bright nebula are strongly anticorrelated to the
surface brightness, even after correcting for pile-up effects.  This
makes a marked contrast to the constancy of the photon index in the
torus, especially considering that the variation of the surface brightness in the
periphery is 1/3 of that in the torus. 

The photon indices of the jet (blue) are generally small compared to
those of the peripheral regions although their surface brightnesses
are comparable.  The jet also shows a strong anticorrelation between
surface brightness and photon index, again suggesting softening of
the underlying electron spectrum at the boundaries of the jet.

The remaining population corresponds to the umbrella region
(yellow). Although distributed over a similar wide range of surface
brightness, the photon indices in the umbrella are generally greater
than those of the torus region. In fact, the umbrella and torus merge
both spatially and spectrally.  The umbrella is seen only at the
northern part of the nebula and has no counterpart with similar
brightness in the south, where the jet is much more distinct.

\section{\label{sec:disc} DISCUSSION}

\subsection{Photon Index Variation}

We found photon index variations across the nebula on arcsecond scales
which are similar to those reported by Weisskopf et al.\ (2000) and
Willingale et al.\ (2001) on larger angular scales. The photon index
variations should be discussed along the two different directions of
particle injection from the pulsar: the direction in the equatorial
plane containing the torus, and the direction of the jet along the
symmetry axis.  We first consider the equatorial plane.

Although the standard model of the Crab Nebula given by Kennel and
Coroniti (1984) assumes a spherical symmetry, it can be adapted to
regions of limited solid angle in the equatorial plane. Across the
equatorial plane, the spectrum is almost constant to the outer
boundary of the torus, but softens significantly beyond this boundary.
In the standard picture, the outer boundary of the torus is
interpreted as a synchrotron burn-off boundary where the synchrotron
losses become significant to the X-ray emitting particles. Until the
outward post-shock flow reaches the torus, the synchrotron critical
energy, $E_c$, must be greater than 10 keV and falls outside the {\it
Chandra} X-ray band (0.5--8 keV). Here, the synchrotron critical
energy is defined as the energy at which a spectral break occurs in
the power-law spectrum.

Due to the shorter synchrotron life time of higher energy particles,
the value of $E_c$ for a given flow element decreases with time, hence
with distance from the shock. Kennel and Coroniti (1984) gave $E_c$ as
a function of radius from the pulsar and explained the energy
dependence of the overall nebular extent from optical to hard X-ray
($\sim$ 60keV). In this picture, if the outer boundary of the X-ray
torus is where the synchrotron burn-off takes place in the X-ray band,
then, in the same manner, larger tori are expected in the optical and
radio bands.  However, it is remarkable that the size of the torus
does not change significantly over a wide bandpass. Optical images
clearly show the torus and related circular structures with the same
size as the X-ray images, albeit with less contrast (Hester et al.\
2002). Temporal variations between two radio images taken 2 months
apart also revealed a torus-like elliptical structure of the same size
(Bietenholz et al.\ 2001). In hard X-ray images (22--43keV and
43--64keV; Pelling et al.\ 1987), again the nebular extent is similar
to that of the torus seen in the soft X-ray band, in contradiction to
an expectation that they should be smaller.  Additionally, Shibata et
al.\ (2003) showed that the observed surface brightness decreases as a
function of distance beyond the torus much faster than expected from
the Kennel \& Coroniti model. Therefore, we suggest that the torus
seen from radio to hard X-rays is not a simple result of synchrotron
burn-off in a radial post-shock flow. The standard Kennel \&
Coroniti flow solution should be modified to include the torus
structure observed in common over the wide bandpass. Detailed
small-scale comparison of the synchrotron burn-off boundary between
the X-ray and optical bands will reveal the true nature of the torus,
which is the subject of a subsequent paper.

Figure~\ref{fig:jet} shows a close-up view of the photon index map
around the southern jet. It shows that the spectral softening occurs
from the central core of the jet, with $\approx \alpha$ of 2.0, to the
outer fainter sheath regions, with $\approx \alpha$ of 2.5.
Considering the cylindrical structure of the jet, the observed
spectrum of the jet core is integrated through this structure and
includes contributions from the softer outer sheath regions. The true
photon index of the jet core is therefore even harder than 2.0, and
must be close to that seen within the torus.  Willingale et al.\
(2001), using {\it XMM-Newton}, reported that the jet has a spectrum
steeper than the torus by 0.3, implying that the corresponding
electron spectra differ by 0.6. However, this conclusion is based on the spectrum
integrated all over the jet. The similarity of the jet core spectrum
to that of the torus indicates that the electron spectra are actually rather
similar in the two different directions of particle injection from the
pulsar. This may give a strong constraint on theoretical models of
magnetic collimation.

We estimate the fraction of the X-ray luminosity that comes from the
jets as $\sim 4$\%, with $\sim$3\% from the southern jet and $\sim$1\%
from the counter jet. The bright umbrella makes it difficult to estimate
the luminosity from the counter jet and results in a relatively large
uncertainty in its luminosity, although the discussion here is little
affected.  The jet luminosity fraction is close to the ratio of the
volumes of the jets and the torus,

\begin{equation}
 \frac{2 \, \int_{0}^{R_{j}} \int_{0}^{\theta_{j}} \int_{0}^{2\pi}
 r^{2}\,\sin\theta\, dr\, d\theta \, d\phi} {\int_{0}^{R_{t}}
 \int_{\frac{\pi}{2} - \theta_{t}}^{\frac{\pi}{2} + \theta_{t}} \int_{0}^{2\pi}
 r^{2} \sin\theta\, dr\, d\theta\, d\phi} =
 \frac{1-\cos\theta_{j}}{\sin\theta_{t}}\left(\frac{R_{j}}{R_{t}}\right)^{3},
\label{eq:ratio}
\end{equation}

\noindent 
where $R_{t},\ R_{j},\ \theta_{t}$ and $\theta_{j}$ are the radii and
semi-opening angles of the torus and the jet. From
Figure~\ref{fig:index_map}a, we see that $R_{t} \approx R_{j}$ and
that $\theta_{t}$ and $\theta_{j}$ are 12$^{\circ}$ and 8$^{\circ}$,
respectively. Then, Equation (\ref{eq:ratio}) gives a ratio of about
5\% for these volumes. Considering the uncertainties in the opening angles and model
assumptions, the fraction of the total X-ray luminosity in the jet is
comparable to its volume fraction, indicating that the volume
emissivities of the jet and the torus are similar, even though they
represent orthogonally different directions of particle injection from
the pulsar. Taking $R_{t}=R_{j}=50$ arcseconds ($\simeq$ 0.5 pc
assuming the distance of 2 kpc to the Crab Nebula) in the above
equation (see Fig.~\ref{fig:index_map}a) and an X-ray luminosity in
the 0.5--8 keV band of 1.3 $\times 10^{37}$ erg s$^{-1}$, which is
estimated from the canonical spectral parameters, we obtain an average
volume emissivity of $\sim$ 4 $\times 10^{-18}$ erg s$^{-1}$
cm$^{-3}$.

\subsection{Surface Brightness Ratio between Near and Far Sides of the Torus}

The north-western (NW) part of the torus, which is the near side, is
much brighter than the south-eastern (SE) part.  The post-shock flows
in the NW and SE parts are relativistically approaching and receding
from us, respectively. Therefore, those apparent surface brightnesses
are subject to Doppler boosting and relativistic
aberration. Considering these effects, the surface brightness ratio of
the NW part to the SE part should be given by

\begin{equation}
 \left(\frac{1+\beta \, \cos\theta}{1-\beta \, \cos\theta}
 \right)^{\alpha+2},
\end{equation}

\noindent 
where $\theta$ is the angle measured in the observer's frame between the
direction of the flow and the line of sight, $\alpha$ is the photon
index, and $\beta$ is the flow speed at the torus normalized to the
speed of light (Rybicki \& Lightman 1979). 

Hester et al.\ (2002) showed that the inner ring, which is inside the
torus, is highly variable and that wisps moving at about 0.5$c$ are
emerging from the inner ring. This provides strong evidence that the
inner ring represents the pulsar wind termination shock, whereas the
torus is a downstream structure. Since the post-shock flow is
sub-sonic, the flow speed decreases as a function of distance. Hester
et al.\ (2002) and Mori (2002) showed that the {\it projected} speed
of the far side of the torus is 0.03$c$--0.1$c$, measured directly
east of the pulsar. The relatively large uncertainty is due to the
small angular displacement of the far side of the torus.  Considering
relativistic effects and the geometry of the torus, where the
rotational-symmetry axis lies about 58$^{\circ}$ west of north and
about 25$^{\circ}$ out of the celestial plane, these speeds correspond
to {\it true} speeds of 0.07$c$--0.25$c$ (0.15$c$). The value in
parentheses is calculated for the central value of the {\it projected}
speed range, 0.065$c$. With $\alpha = 1.9$, these values should result
in a surface brightness ratio of 1.6--6.3 (2.9).  The observed surface
brightness ratio in Figure~\ref{fig:index_map}a is about 3.4, which
agrees with the speed range inferred from above discussion.

Kennel \& Coroniti (1984) deduced the post-shock flow speed as a
function of $z$, the distance from the pulsar normalized to the shock
radius, for different values of the magnetization parameter $\sigma$,
the ratio of the Poynting flux to the particle kinetic flux upstream
of the shock.  From Figure~\ref{fig:index_map}a, $z$ is about 2.7 at
the distance where we measured the speed of the torus.  Kennel \&
Coroniti suggested that $\sigma$ is about 0.003 based on theoretical
arguments, which would correspond to a velocity of about 0.05$c$ at
the torus (from Fig.~3 of Kennel \& Coroniti), far below our measured
value.  Shibata et al.\ (2003) constructed a two-dimensional model
image of the Crab Nebula in the framework of the Kennel \& Coroniti
model, and found that the model surface brightness ratio between the
near and far sides of the torus is about 1.3 for $\sigma = 0.003$,
also much lower than the observed value.  By contrast, our measured
speed of $0.07c$--$0.25c$ (0.15$c$), which predicts the correct
surface brightness ratio, corresponds to $\sigma = 0.01$--0.13 (0.05).
The theoretical speed and brightness ratio values can be reconciled
with the measured ones if we assume a $\sigma$ of about 0.05.

It is worth noting that departures from spherical symmetry and proper
treatment of radiative losses in the flow can both affect this result.
It is likely that the flow is confined to a narrow range of latitudes
around the equatorial plane upstream of the shock (the inner ring).
After the shock the flow diverges out of the plane more rapidly than a
radial flow to form the broad torus. This should result in more rapid
deceleration than in a spherical model.  Any radiative losses of
particle pressure may also lead to greater compression and hence more
rapid deceleration.  In both cases, however, we might expect flow
velocities {\it slower} than those predicted by Kennel \& Coroniti,
for a given $\sigma$.  As a result, either of these effects could
require an even greater $\sigma$ to account for the observations.

\subsection{\label{sec:nh} Correlation between Column Density and Ejecta}

\begin{figure}[]
\epsscale{1.0}
\plotone{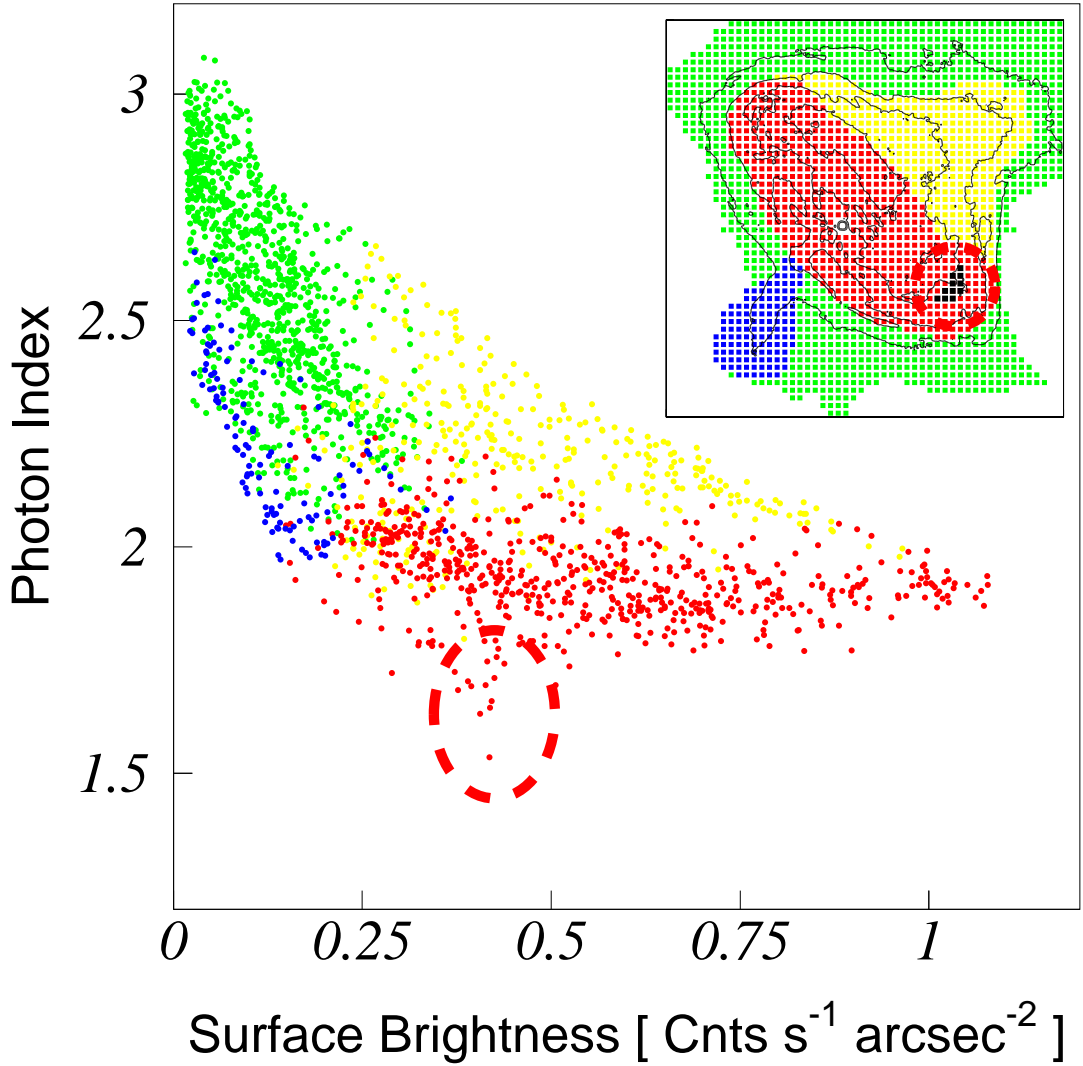}
\caption{Plot of photon index against surface brightness, same as
 Figure~\ref{fig:correlation_plot}b. The data points of the torus, the
 jet, the umbrella-shaped northwest region, and the faint peripheral
 region are color-coded as red, blue, yellow, and green, respectively.
 The inset shows the definitions of the four regions. The black
 contours represent surface brightness. The red dashed circle encloses
 data points with unusually small photon index compared to others (see
 \S \ref{sec:nh}). The corresponding positions in the nebula are shown
 in black in the inset, also enclosed by a red dashed circle.
\label{fig:color_plot}}
\end{figure}

\begin{figure}[]
\epsscale{1.0}
\plotone{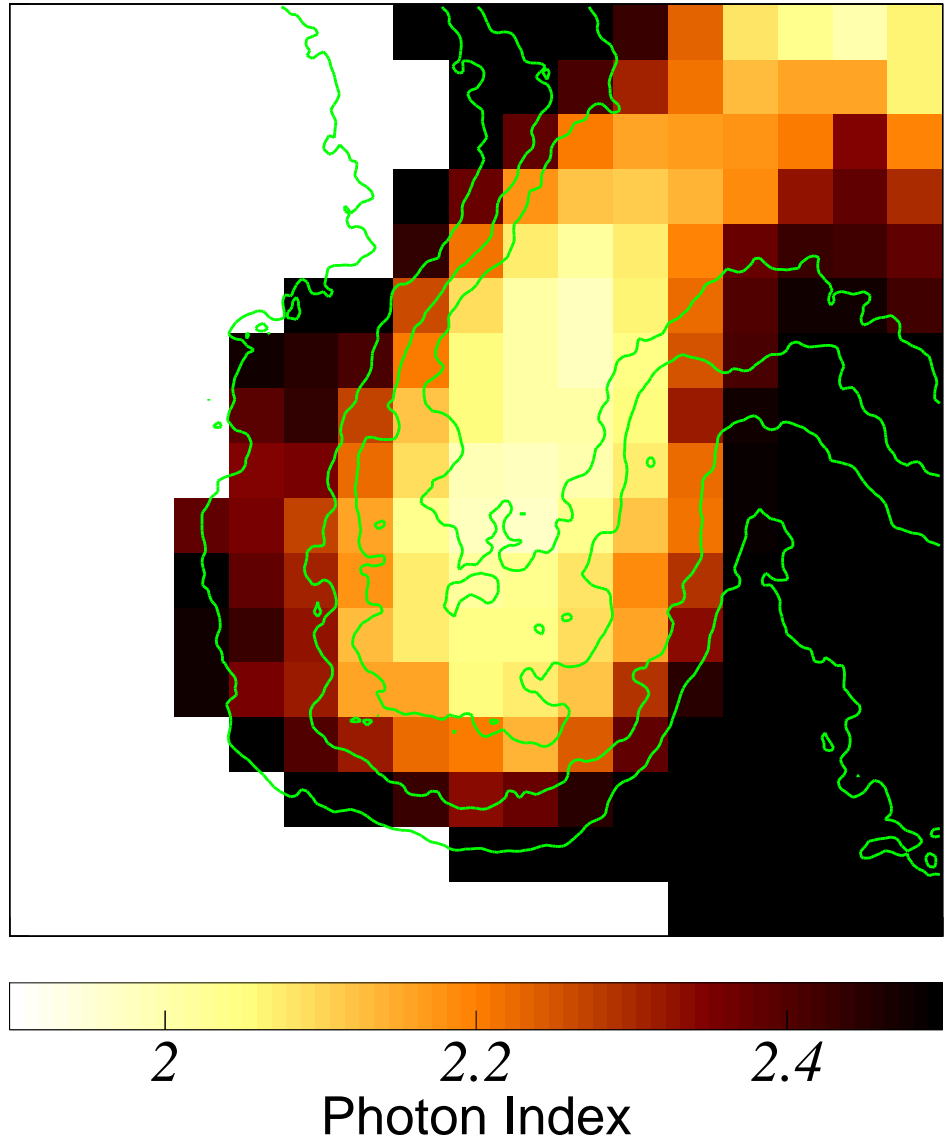}
\caption{Close-up view of photon index map around the southern jet.
 Note that the range of photon index is different from that in
 Figure~\ref{fig:index_map}a to show fine variation. The green
 contours represent surface brightness.
\label{fig:jet}}
\end{figure}

\begin{figure}[]
\epsscale{0.8}
\plotone{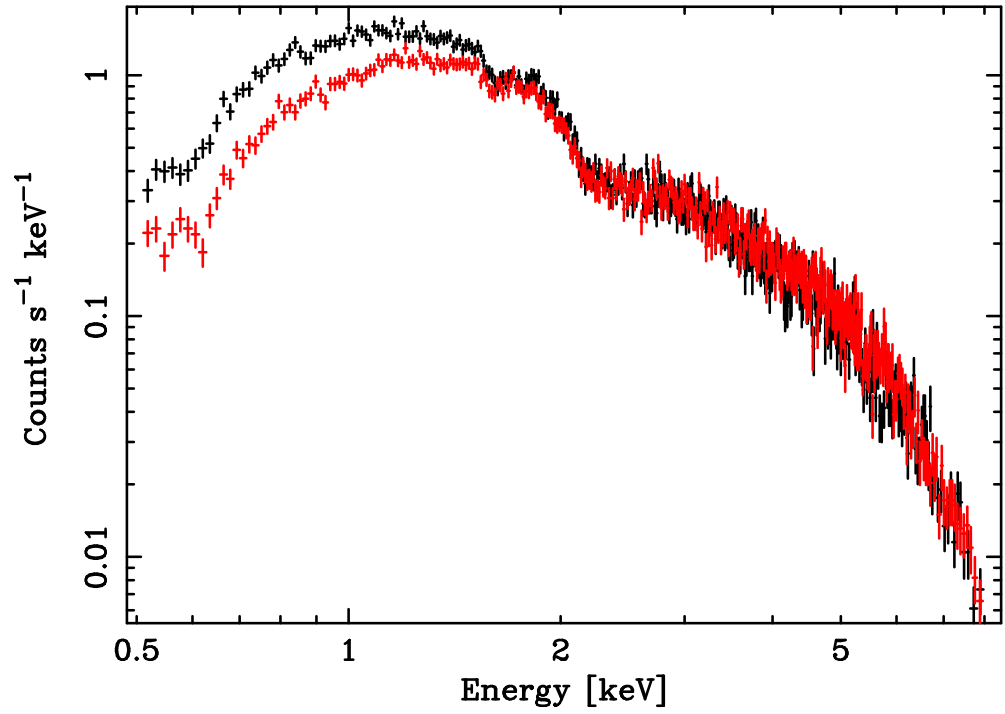}
\caption{Spectra extracted from one of the small photon index
regions (red) and a neighboring region (black).
\label{fig:spectrum_comparison}}
\end{figure}

There are some regions exhibiting unusually small photon index in
Figure~\ref{fig:color_plot}, which are enclosed by the red dashed
circle.  The corresponding positions in the nebula are shown in black
in the inset, also enclosed by a red dashed
circle. Figure~\ref{fig:spectrum_comparison} presents two spectra
extracted from one of the small photon index regions and its
neighboring region. The spectra are identical above about 2 keV, and
the comparison of the spectra clearly reveals that the small apparent
photon index is actually caused by a deficit of soft emission due to
increased absorption in this region.  Refitting the spectra from those regions with
$N_{\rm H}$ treated as a free parameter, we obtained a similar photon index and a
higher $N_{\rm H}$, by $\sim 7 \times 10^{20}$ cm$^{-2}$, compared to
the neighboring region.

The Crab Nebula appears to contain a number of embedded filaments in
optical narrow-band emission-line images (e.g., Sankrit \& Hester
1997). The optical filaments are thought to be the result of a
Rayleigh-Taylor instability between the light pulsar wind nebula and
the dense ejecta of the supernova explosion (Hester et al.\ 1996).
The J2000 position of the region showing higher $N_{\rm H}$, $\alpha=$
05$^{\mathrm h}$34$^{\mathrm m}$29\fs6,
$\delta=+$22\arcdeg00\arcmin30\farcs4, coincides with knotty structure
in the most prominent optical filament, which is often referred to as
part of the ``High-Helium Band'' (MacAlpine et al.\ 1989). The
emission lines from the High-Helium Band are blueshifted, indicating that this
filament is located between the synchrotron nebula and us. In an
optical continuum image, the knotty structure is visible as a
``shadow'' because it consists of dense dust which absorbs the optical
synchrotron continuum from the PWN (e.g., Blair et al.\
1997). Similarly, the higher $N_{\rm H}$ we find is likely due to
absorption of soft X-rays by the dust and gas in this knotty
structure. Assuming that the thickness of this filament is
5$^{\prime\prime}$ based on the optical emission line images (Blair et
al.\ 1997), which corresponds to $\sim 1.5 \times 10^{17}$ cm, the
density of this filament corresponding to the observed X-ray
absorption column density is $\sim 5 \times 10^{3}$ cm$^{-3}$. The
excess absorption can be seen only around the knotty structures all
through this filament.

We searched for other examples of X-ray absorption from the ``shadow'' features in the optical
continuum image (Hester et al.\ in preparation), from which we can
pick up only dense dust core in filaments on the near side of the
nebula. There are only a few such shadows within the X-ray
nebula. The number of corresponding analysis regions is about 10 out of
2074 (in addition to the regions discussed above). The second noticeable
shadows are located to the east of the pulsar, overlapping the edge of the
torus. Excess absorption was again observed, but the amount was
$\sim 3 \times 10^{20}$ cm$^{-2}$. The excess absorption related to
other shadows is similar or less. Therefore, our results are little
affected by soft X-ray absorption due to the filaments.

It is conceivable that our results are affected by intrinsic column
density variation due to not only the optical filaments but also
arcsecond scale variation of galactic absorption. We assessed this
effect as follows. In Figure~\ref{fig:color_plot}, the photon indices
of the data points of the torus (red) having surface brightness higher
than 0.5 counts s$^{-1}$ arcsec$^{-2}$ are distributed around the mean
of 1.91 with the root mean square of 0.07. The statistical errors are
small, so the observed scatter is dominated by systematic errors and
by real variations in the power law index.  We can obtain an upper
limit on possible absorption variations across the nebula if we assume
that this scatter is produced by variations in $N_{\rm H}$, leading to
an estimate of the variation in $N_{\rm H}$ of $2 \times 10^{20}$
cm$^{-2}$.

\section{\label{sec:summary} SUMMARY}

We have shown spatial variations of the X-ray spectrum of the Crab
Nebula in terms of photon index, at an angular scale of
arcseconds. The variations can be viewed in two different directions
of the particle injection from the pulsar.

Across the equatorial plane, the spectrum is almost constant to the
outer boundary of the torus, with photon index $\alpha$ $\approx$ 1.9
regardless of the surface brightness.  It softens significantly, up to
$\alpha$ $\approx$ 3.0, in the outer, fainter peripheral region.  This
seems qualitatively consistent with the previous suggestions that the outer boundary
of the torus is interpreted as a synchrotron burn-off boundary where
the synchrotron losses become significant to X-ray emitting
particles. However, the fact that structures similar to the torus are
seen at other wavelengths indicates that the torus is not a simple
result of synchrotron burn-off.

Within the southern jet, photon index variations are also seen: the
spectral softening takes place from the central core to the outer
sheath. The photon index at the central core is almost the same as
that of the torus. This indicates that the electron spectra are
similar in the two different directions of the particle injection from
the pulsar. We also found that the volume emissivities of the jet and
the torus are similar.

Assuming that the brightness difference between the near and far sides
of the torus is caused by Doppler boosting and relativistic
aberration, the ratio can be explained by the observed speed of the
downstream flow at the torus. However it cannot be obtained from the
so-called weakly magnetized pulsar wind with $\sigma \approx 0.003$,
as suggested by Kennel \& Coroniti (1984).

Finally, we found that an optical filament comprised of supernova ejecta
surrounding the pulsar wind nebula is absorbing the soft X-ray
emission from a small portion of the X-ray nebula.

\acknowledgments

The authors thank G.\ Chartas for his extensive cooperation on
simulation using LYNX. We also thank the anonymous referee for comments
which significantly improved the final text of this paper. K.\ M.\
acknowledges the support of JSPS through the fellowship for research
abroad. The work of G.\ G.\ P.\ was partially supported by NASA grant
NAG5-10865.  This work was supported by the NASA through Chandra Awards
G00-1032B and G01-2076B, issued by the Chandra X-ray Observatory Center,
which is operated by the Smithsonian Astrophysical Observatory for and
on behalf of NASA under contract NAS8-39073.

\clearpage

\begin{deluxetable}{cccc}
\footnotesize
\tablecaption{Observational log of the Crab Nebula \label{tbl:log}}
\tablewidth{0pt}
\tablehead{ \colhead{ObsID} & \colhead{Date} & 
 \colhead{Coord.(J2000)\tablenotemark{a}} & 
 \colhead{Exposure (sec)} }
\startdata
 1994 & 3 November 2000 & 83.637, 22.024  & 3241.8 \\
 1996 & 25 November 2000 & 83.632, 22.014  & 2926.8 \\
 1998 & 18 December 2000 & 83.631, 22.016  & 2703.8 \\
 1999 & 9 January 2001 & 83.631, 22.017  & 2663.4 \\
 2001 & 30 January 2001 & 83.631, 22.017  & 2553.0 \\
 1995 & 21 February 2001 & 83.631, 22.017  & 2648.0 \\
 1997 & 14 March 2001 & 83.631, 22.017  & 2568.6 \\
 2000 & 6 April 2001 & 83.631, 22.016  & 2643.0 \\
\enddata
 \tablenotetext{a}{coordinates of the center of the subarray}
\end{deluxetable}

\clearpage

\begin{deluxetable}{lccc}
\rotate
\footnotesize
\tablecaption{Spectral parameters for whole nebula \label{tbl:whole_fit}}
\tablewidth{0pt}
\tablehead{ \colhead{Parameter} & \colhead{Uncorrected Value} &
\colhead{Corrected Value} & \colhead{Canonical Value} }
\startdata
$N_{\rm H}$ (10$^{22}$ cm$^{-2}$)  & 0.27  & 0.32 (fixed) & 0.33--0.36\tablenotemark{a} \\
Photon index          & 1.92  & 2.10 & 2.10 $\pm$ 0.03\tablenotemark{b}        \\
Normalization\tablenotemark{c}  & 6.57  & 9.65  & 9.7 $\pm$ 0.5\tablenotemark{d} \\
\enddata
 \tablenotetext{a}{Kuiper et al.\ (2001)}
 \tablenotetext{b}{Toor \& Seward (1974)}
 \tablenotetext{c}{(Photons cm$^{-2}$ s$^{-1}$ keV$^{-1}$ at 1 keV)}
 \tablenotetext{d}{Willingale et al.\ (2001)}
\end{deluxetable}

\end{document}